\documentclass[aps,prl,twocolumn,groupedaddress,showpacs]{revtex4}
\usepackage{graphicx}
\usepackage{dcolumn}
\usepackage{bm}
\usepackage{amsmath}
\usepackage{color}
\begin{document}
\title{A pulsed Sisyphus scheme for laser cooling of atomic (anti)hydrogen}
\date{\today}
\author{Saijun Wu}
\author{Roger C. Brown}
\author{William D. Phillips}
\author{J. V. Porto}
\affiliation{Joint Quantum Institute, NIST and University of Maryland,
Gaithersburg, Maryland 20899}
\begin{abstract}
We propose a laser cooling technique in which atoms are selectively excited to a dressed metastable state whose light shift and decay rate are spatially correlated for Sisyphus cooling. The case of cooling magnetically trapped (anti)hydrogen with the 1S-2S-3P transitions using pulsed ultra violet and continuous-wave visible lasers is numerically simulated. We find a number of appealing features including rapid 3-dimensional cooling from $\sim$1~K to recoil-limited, millikelvin temperatures, as well as suppressed spin-flip loss and manageable photoionization loss.
\end{abstract}
\pacs{37.10.De, 67.63.Gh}

\maketitle

Recent progress~\cite{octupoletrap,cusptrap} in producing antihydrogen ($\overline {\rm H}$) improves the prospects for precision spectroscopy and de Broglie wave interferometry of $\overline {\rm H}$ that may uncover new physics in low-energy experiments~\cite{athena, atrap}. Antihydrogen atoms are synthesized from anti-protons and positrons~\cite{athena, atrap, penningtrap, octupoletrap,cusptrap} in such small numbers that trapping and efficient cooling to millikelvin temperatures are likely required for precise measurements. In contrast to hydrogen, which may be cooled by collisions with a buffer gas~\cite{buffergas05}, or by selecting low-energy atoms from an intense beam, cooling of $\overline {\rm H}$ will likely rely on laser cooling techniques~\cite{metcalf}.


The most obvious approach to laser cooling of ${\rm H}$~\cite{Setija93} or $\overline {\rm H}$ demands 121.6~nm Ly-$\alpha$ radiation. Apart from the difficulty of manipulating VUV light, the generation of even $\sim$10~nW of CW Ly-$\alpha$ radiation is technically challenging~\cite{Scheid09}. In addition, such cooling of $\overline {\rm H}$ atoms in a kelvin-deep magnetic trap faces several interrelated difficulties. The need to avoid spin-flip losses, combined with the limited fraction of phase space addressable with a low intensity, single-frequency laser, implies that the cooling will be slow. Indeed, the only experimental Ly-$\alpha$ cooling work so far~\cite{Setija93} used a pulsed laser with an average power of 160 nW (2.5~nW at the location of atoms) to cool magnetically trapped ${\rm H}$, and took more than 15 minutes to reach 8 mK, starting from just 80~mK. Furthermore, 3D cooling in this approach was aided by collisional mixing, which will be absent in dilute samples of ${\overline {\rm H}}$. Instead of relying on Ly-$\alpha$ radiation, there are several proposed cooling schemes using more readily available lasers to drive Doppler sensitive 2-photon transitions~\cite{Allegrini93,Zehnle01,ultrafast06}. However, in addition to limited phase-space addressability similar to Ly-$\alpha$ cooling, these schemes have the difficulty of losses due to photoionization.




\begin{figure}
\includegraphics [width=3.2 in,angle=0] {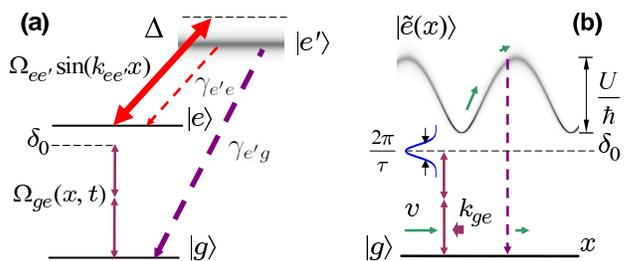}
\caption{(a): Level diagram for the proposed cooling scheme (see text). (b): Simplified dressed-state picture of the cooling scheme. The pulsed two-photon excitation has bandwidth $1/\tau$, is detuned from the bottom of the lattice with depth $U$ by $\delta_0$, and provides Doppler cooling by transferring momentum $\hbar k_{ge}$ to an atom with velocity $v$. The linewidth $\Gamma(x)$ of the dressed excited state $|\tilde e(x)\rangle$ is indicated by the width of the gray curve.  As the atom climbs the hill the velocity (green arrow) decreases while the decay probability increases, leading to Sisyphus cooling. (The $|e'\rangle\rightarrow |e\rangle$ decay is ignored in (b). Not shown is the other, detuned dressed state $|\tilde e'(x)\rangle$.) }\label{figScheme}
\end{figure}


Motivated by previous work~\cite{multiphoton09}, which used transitions between excited states for cooling and trapping, we propose a 3-level cooling scheme (Fig.~\ref{figScheme}), applicable to magnetically trapped H, where a metastable state $|e\rangle$ is coupled to a short-lived state $|e'\rangle$ by a blue detuned standing wave coupling $\Omega_{ee'}$. Atoms in the ground state $|g\rangle$ are repeatedly excited to the bottom of the dissipative $|e\rangle-|e'\rangle$ optical lattice by a pulsed, Doppler sensitive 2-photon coupling $\Omega_{ge}$. The proposed cooling process arises from two effects: 2-photon Doppler cooling~\cite{Allegrini93,Zehnle01,ultrafast06} associated with $|g\rangle \rightarrow |e\rangle$ excitation, and Sisyphus cooling~\cite{aspect86} associated with the $|e\rangle-|e'\rangle$ lattice.

For magnetically trapped H, $|g\rangle$, $|e\rangle$ and $|e'\rangle$ are the maximally Zeeman shifted states in the 1S, 2S and 3P manifolds respectively.  We show that the cooling scheme provides both a large capture velocity (100 m/s) and a low final temperature (near the Ly-$\beta$ single-photon recoil temperature of 1.8~mK), and allows for large volume 3D cooling. Advantages include: availability of both nanosecond-pulsed UV 2-photon 1S-2S radiation~\cite{Yatsenko99} and CW 2S-3P radiation at 656 nm; reduction of UV photoionization losses; and suppressed spin-flip transition from the 3P level in a high field. In the following we first discuss the pulsed Sisyphus cooling scheme in a 1D, semiclassical model. After justifying the model with a 1D quantum simulation~\cite{Dalibard92,Camichael93}, we present a 3D semiclassical simulation for magnetically trapped ${\rm H}$.


The proposed cooling scheme involves repeated pulsed excitations, each followed by spontaneous decay~\cite{metcalf2}. The propagation direction of the 2-photon excitation pulses alternates between $\pm \hat x$. The Rabi frequency $\Omega_{ge}(x,t)$ is $(\theta /\tau) f(t/\tau)e^{\pm i k_{ge} x}$, where $\theta$ is the pulse area, $k_{ge}$ is the sum of the wavevectors of the two photons, and $f(t/\tau)$ is a normalized pulse-shape function with characteristic duration $\tau$. For $\theta\ll 1$ its Fourier transform $F(\omega\tau)$ gives the excitation spectrum. The interval $T_{\rm rep}$ between pulses is long enough so that excited atoms, moving in the $|e\rangle-|e'\rangle$ lattice, decay to $|g\rangle$ with high probability. In the effective 2-level system (Fig.~\ref{figScheme}b), the spatially dependent detuning $\delta(x)$ and linewidth $\Gamma(x)$ (see Eq.~(\ref{eq_adpotential})) allows spatial selectivity in both pulsed excitation and subsequent decay. The $|e\rangle - |e'\rangle$ transition is driven by a standing wave coupling $\Omega_{ee'}(x)$ with a positive detuning $\Delta$, resulting in two dressed states $|\tilde e(x)\rangle$ and $|\tilde e'(x)\rangle$ which are spatially dependent superpositions of $|e\rangle$ and $|e'\rangle$ and connect to those states respectively as $\Omega_{ee'}\rightarrow 0$. The 2-photon $|g\rangle-|e\rangle$ detuning from the unshifted metastable state $|e\rangle$ (decay rate $\gamma_{e g}\approx 0$) is $\delta_0$, and $\gamma_{e'g},\gamma_{e'e}$ are the decay rates from $|e'\rangle$ (Fig.~\ref{figScheme}a). We assume $\Delta\gg \delta_0,\gamma_{e'g}$ and $\gamma_{e'g}\gg \gamma_{e'e}$. Atoms are predominantly excited to, and adiabatically follow, the dressed state $|\tilde e(x)\rangle$ (and not $|\tilde e'(x)\rangle$)~\cite{exp:foot1}, and
\begin{equation}
\begin{array}{l}
\delta(x)= \delta_0-(\sqrt{\Omega_{ee'}(x)^2+\Delta^2}-\Delta)/2, \\
\Gamma(x)= \frac{\delta_0-\delta(x)}{\sqrt{\Omega_{ee'}(x)^2+\Delta^2}} \gamma_{e'g}.
\end{array}\label{eq_adpotential}
\end{equation}
In Eq.~(\ref{eq_adpotential}) we have ignored $\gamma_{e'e}$ and $\gamma_{e g}$, but their inclusion has little influence on the results described below. The depth $U$ of the resulting $|e\rangle-|e'\rangle$ optical lattice is given by the maximum of $\delta_0-\delta(x)$, and its period is determined by $k_{ee'}$. We define $k_{i j}$ as the wavevector of the $i-j$ transition for $i,j=g,e,e'$, with the associated recoil velocities and frequencies defined as $v_{i j}=\hbar k_{i j}/m$ and $\omega_{r,i j}=\hbar k_{i j}^2/(2 m)$  where $m$ is the atomic mass.


Doppler-sensitive $|g\rangle-|e\rangle$ absorption leads to 2-photon Doppler cooling~\cite{Allegrini93,Zehnle01,ultrafast06}. The Doppler shifted, spatially dependent detuning, $\delta(x,v)=\delta(x)\pm k_{ge}v$, allows atoms at different velocities to be excited to the lattice potential at different locations.  Atoms with velocity $v$ are resonantly excited at positions such that $|\delta(x, v)|\tau \lesssim 1$. For $\tau>2\pi/|\delta_0|$, strong excitation only occurs in the velocity range $v_d < |v| < v_c$, with  decoupling velocity $v_d \simeq |\delta_0| / k_{ge}$ and capture velocity $v_c \simeq (|\delta_0|+U/\hbar)/k_{ge}$. Atoms with $|v| \ll v_d$ or $|v| \gg v_c$ are off resonance and not efficiently excited.


In addition to Doppler cooling, the correlation between the spatially dependent detuning $\delta(x)$ and the decay rate $\Gamma(x)$ leads to Sisyphus cooling since atoms preferentially decay from $|\tilde e(x)\rangle$ at the tops of the light shift potential~\cite{aspect86}. The Sisyphus effect is particularly efficient for atoms with $v$ close to $v_d$, which are excited near the bottom of the lattice. If $\frac{1}{2} m v^2<U$, atoms remain within one lattice site and typically oscillate before decaying to $|g\rangle$. The decay is enhanced at the classical turning point, due to both the larger decay rate and the longer time spent there. Averaged over the position (and velocity) dependent decay probability, the velocity distribution after decay is centered at zero velocity, with a rms width less than $\frac{1}{2} v$ for $v\gg v_{g e}$. On average this removes more than 75$\%$ of the atomic kinetic energy per 2-photon excitation.

\begin{figure}
\includegraphics [width=2.8 in,angle=0] {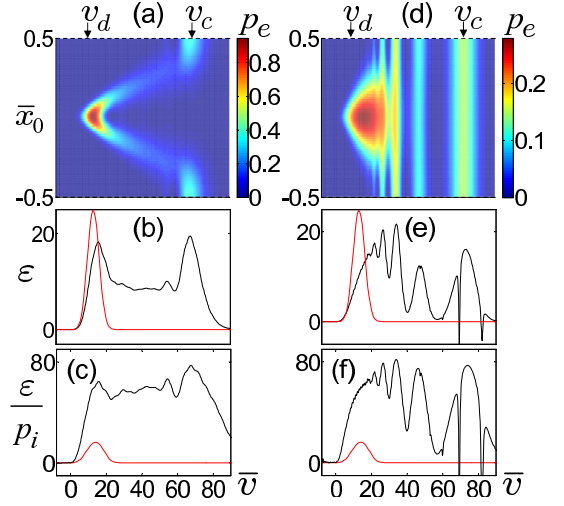}
\caption{OBE simulation of cooling properties. (a,d): Normalized excitation probability $p_e$ vs $\bar x_0=k_{ee'} x_0/\pi$ and $\bar v=v/v_{g e}$. (b,e): Normalized energy loss per pulse $\varepsilon$ vs $\bar v$. $\varepsilon$ is averaged over $x_0$ and is in units of $\hbar \omega_{r,ge}$. (c,f): Ratio of $\varepsilon$ to the normalized ionization probability, $\varepsilon/p_i$ vs $\bar v$. Here $\delta_0=-25\omega_{r,ge}$, $\tau=-2.5/\delta_0$,  $f(t)=\frac{1}{\sqrt{\pi}}e^{-t^2}$, $\theta=\pi/8$, $\Omega_{ee'}(x)=\Omega_{ee'}\sin(k_{ee'}x)$, $\Delta=\Omega_{ee'}/2=200~\omega_{r,ge}$, and $\gamma_{e'g}=8\gamma_{e'e}=2\omega_{r,ge}$.  The red curves in (b,c,e,f) correspond to $\Omega_{ee'}=0$. The left and right panels are for $k_{ge}/k_{ee'}$=25 and 5.4 respectively.
}\label{figOBE}
\end{figure}

To characterize the cooling, we define the normalized position and velocity dependent excitation probabilities $p_e = 4P_e/\theta^2$ and energy loss per 2-photon pulse $\varepsilon = 4\Delta E/\theta^2$, where $P_e$ is the 2-photon excitation probability, $\Delta E$ is the energy loss per 2-photon pulse and $p_e$ and $\varepsilon$ are $\theta$-independent for $\theta \ll 1$.
Figure 2(a,b,d,e) shows $p_e$ and $\varepsilon$, determined using 3-level
optical Bloch equations (OBE) for a ``dragged atom''
following a trajectory $x(t) = x_0 + v t$. Figure~\ref{figOBE}(a, b) and Fig.~\ref{figOBE}(d, e) are for $k_{g e}/k_{e e'}=25$ and 5.4 respectively. The latter ratio corresponds to $k_{\rm 1S-2S}/k_{\rm 2S-3P}$ in hydrogen. Figure~\ref{figOBE} suggests that the simple picture of phase-space selective excitation plus Sisyphus cooling only applies for $k_{ee'}\ll k_{ge}, (v \tau)^{-1}$, as in Fig.~\ref{figOBE}(a,b). For atoms that move more than $1/k_{ee'}$ during $\tau$, the excitation is complicated by multi-photon resonances at velocities with $k_{ge} v +2 n k_{ee'}v \approx \delta$ (the peaks of the black curve in Fig.~\ref{figOBE}e for $\bar v>20$)~\cite{multiphoton09}. In addition, for large $k_{ee'}$, Doppleron resonant coupling to $|\tilde e'(x)\rangle$~\cite{Doppleron90} occurs at moderate speeds $v$ with $2 n k_{ee'}v \approx \Delta $ for integer $n$ (the sharp dips of the black curve in Fig.~\ref{figOBE}e near $ \bar v=70,~80$), and leads to heating. Nevertheless, efficient Sisyphus cooling is still possible for moderate $k_{g e}/k_{e e'}$ (see Fig.~\ref{figOBE}e). Compared to regular 2-photon cooling (red curves in Fig.~\ref{figOBE}(b,e)), the peak excitation probability is decreased by approximately $\frac{2\pi}{\tau}/\frac{U}{\hbar}$ due to the spatially inhomogeneous broadening of $|\tilde e(x)\rangle$ (Fig.~\ref{figScheme}b). However, due to the Sisyphus enhanced energy removal per excitation, the average energy loss per pulse remains comparable to the Doppler-only case, but with an increased velocity capture range $v_d<|v|<v_c$.

In addition to the increased velocity capture range, the decreased excitation probability to $|\tilde e(x)\rangle$ also helps mitigate the photonionization loss from $|\tilde e(x)\rangle$ to the continuum. For degenerate 2-photon excitation of H to the 2S level, the ionization probability per pulse is given by $P_{\rm ioni}=\int{dt \gamma_{\rm ioni}(t)\rho_{\rm 2S}(t)}$ where $\rho_{\rm 2S}(t)$ is the 2S state population and $\gamma_{\rm ioni}$ is the rate of ionization from 2S due to the UV radiation~\cite{Haas06}. As a measure of cooling efficiency per 2-photon pulse, in Fig.~\ref{figOBE}(c,f) we compare $\varepsilon/p_i$, the ratio between the normalized energy loss $\varepsilon$ and normalized ionization probability $p_i=12 P_{\rm ioni}/\theta^3$, with (black curve) and without (red curve) the Sisyphus cooling. Here we have set $\gamma_{\rm ioni}=1.6 \Omega_{ge}$~\cite{Haas06}.  We see that the Sisyphus effect enhances the cooling efficiency by approximately $U/|\hbar \delta_0|$ near $v=v_d$ where the Doppler cooling has the best $\varepsilon/p_i$.




We simulate the cooling process with a semiclassical stochastic wavefunction (SCSW) method~\cite{Dalibard92,Camichael93}. The simulation of a cooling cycle is divided into two stages: excitation ($0<t<\tau$) and decay ($\tau<t<T_{\rm rep}$) (we ignore quantum jumps during excitation).  The external motion of the atom is described by a classical trajectory $x(t)$. The internal dynamics are described by a stochastic wavefunction $|\psi(t)\rangle$, which, after the $g-e$ pulse, is probabilistically projected to either $|g\rangle$ or the $\{|e\rangle,|e'\rangle\}$ manifold (typically almost all in $|\tilde e(x)\rangle$) as $|\psi_p\rangle$. Due to this post-selection, the optical force in the excitation stage cannot be evaluated in the usual way as $\langle \psi(t) |\hat F |\psi(t) \rangle$ where $\hat F$ is the force operator. Instead, the force is estimated as the real part of a ``weak value''~\cite{Aharonov88}, $\langle \psi_p(t) |\hat F |\psi(t) \rangle/\langle \psi_p(t) |\psi(t) \rangle$, where $|\psi(t)\rangle$ is found by forward-propagating the pre-determined state $|\psi(0)\rangle=|g\rangle$ and $\langle \psi_p(t)|$ is found by back-propagating the post-determined state $\langle \psi_p(\tau)|$, both for a dragged atom. This estimation method reproduces the quantum-mechanically expected velocity change during the pulse, $\delta v_{\rm pulse}$, due to both the recoil effect and the excited-state dipole force. During the second stage, the stochastic wavefunction $|\psi_p(t)\rangle\in \{|e\rangle,|e'\rangle\}$ manifold and $x(t)$ are propagated in small time-steps, until a quantum jump occurs~\cite{Dalibard92,Camichael93}. If the quantum jump is an $|e'\rangle \rightarrow |e\rangle$ transition, we project $|\psi_p(t)\rangle$ to the dressed states $|\tilde e(x)\rangle$ or  $|\tilde e'(x)\rangle$  probabilistically~\cite{ChenJian93}, while for an $|e'\rangle \rightarrow |g\rangle$ jump, we propagate $x(t)$ freely until the next pulse. Upon each spontaneous emission, we use random velocity jumps to account for the recoil effect.


\begin{figure}
\includegraphics [width=3. in,angle=0] {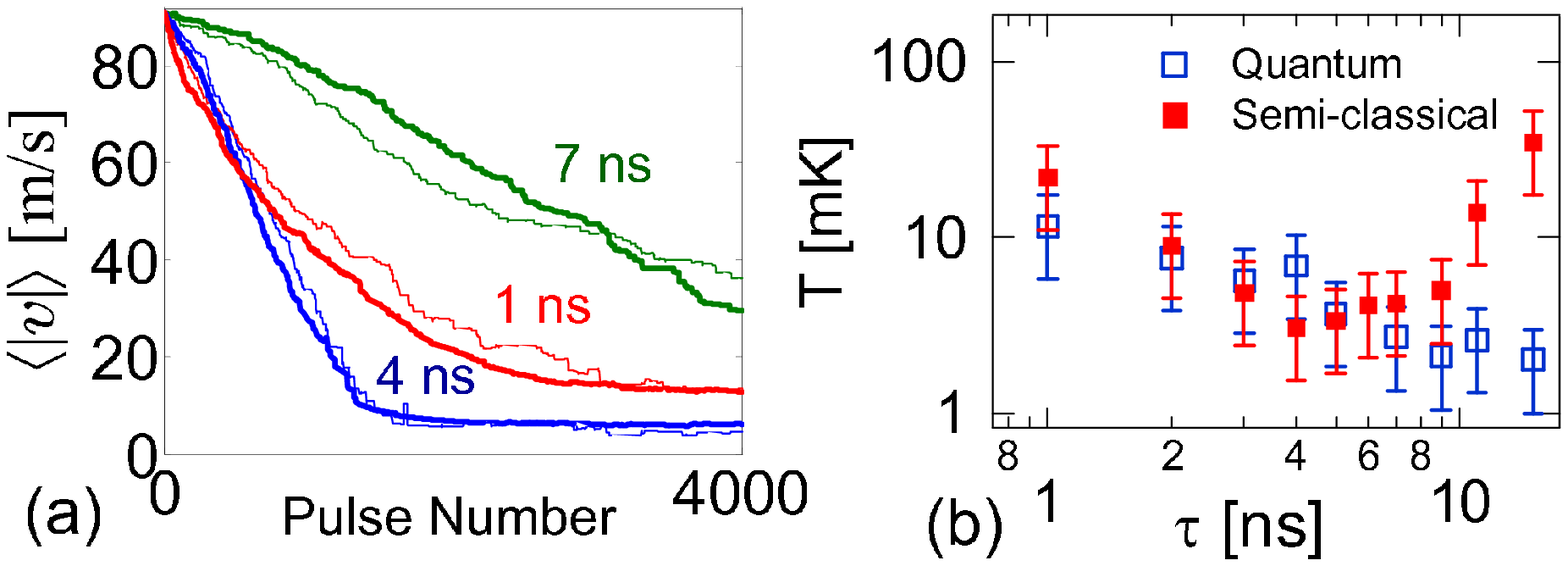}
\caption{Comparison of 1D SCSW and QSW simulations. Here $\Omega_{ee'}/(4\pi)=\Delta/(2\pi)$=2.7 GHz as in Fig.~\ref{figOBE}, $\delta_0=-2\pi/\tau$,  $\theta=\pi/4$. (a): Average speed $\langle|v|\rangle$ vs pulse number for three different pulse durations $\tau$. 4~ns represents a compromise between 1~ns, where the bandwidth is so large that there is little spatial selectivity, and 7~ns, where the bandwidth is so small that only a small fraction of atoms are excited. Thick lines are an average of 30 QSW trajectories, while thin lines are an average of 20 SCSW trajectories. (b): The equilibrium temperature ${\rm T}$ vs pulse duration $\tau$. At small velocities the SCSW method becomes less accurate for $\tau>9$~ns, roughly set by half the oscillation period in a single $|e\rangle-|e'\rangle$ lattice site.
}\label{FigCQ}
\end{figure}

We use a 1D full quantum stochastic wavefunction (QSW) simulation~\cite{Dalibard92,Camichael93}, which includes both internal and external degrees of freedom of a 3-level atom, to confirm that the SCSW method correctly predicts the cooling dynamics and the final temperature. In Fig.~\ref{FigCQ} typical results for smoothed square pulses~\cite{exp:foot5} are compared, for the appropriate hydrogen 1S-2S-3P parameters, $\gamma_{e'g}/(2\pi) = 26.6$~MHz, $\gamma_{e'e}/(2\pi)= 3.6$~MHz,  $\{v_{g e}, v_{e e'},v_{g e'}\}=\{3.3,0.6,3.9\}$~m/s, and $\{\omega_{r,ge}, \omega_{r,ee'},\omega_{r,ge'}\}/(2\pi)=\{13.4,0.46,18.8\}$~MHz, which are also used in Fig.~\ref{figOBE}(d-f). We find good agreement between the SCSW and QSW methods as long as the dragged atom picture is valid during the pulse, i.e., if the optical force during the short excitation does not significantly displace the trajectory compared to the wavelength ($k_{ee'}~\delta v_{\rm pulse}~\tau\ll 1$)(Fig.~\ref{FigCQ}b). The attainable 1D temperature predicted by the quantum simulation decreases with decreasing bandwidth $1/\tau$, and is remarkably low ($\sim$3 mK) even with $\tau=5$~ns.

Having verified the semiclassical approach for our parameters, we use SCSW to simulate 3D cooling of magnetically trapped H. To describe the 3D light-atom interaction, we included all ten electronic levels in the 1S-2S-3P manifold (ignoring hyperfine structure). In a high magnetic field, the cooling process is dominated by the three maximally Zeeman-shifted states of the 1S, 2S and 3P levels (corresponding to $|g\rangle$, $|e\rangle$ and $|e'\rangle$), which would form a closed system under 1S-2S 2-photon coupling $\Omega_{ge}$ and perfect $\sigma^+$ 2S-3P coupling $\Omega_{ee'}$. We consider a magnetic trap with ${\bf B}=\{B_x,B_y,B_z\}=\{B_1 y-B_2 z x/2,B_1 x-B_2 z y/2, B_0+B_2 (2 z^2 -x^2-y^2)/4\}$, $B_0=0.75$~T, $B_1$=0.8~T/cm and $B_2=12$~mT/cm$^2$, similar to those for an existing antihydrogen apparatus~\cite{penningtrap}. Both $|g\rangle$ and $|e\rangle$ feel a trapping potential $V\approx \mu_{\rm B}|{\bf B}|$ ($\mu_{\rm B}$ is the Bohr magneton), so the $|g\rangle-|e\rangle$ detuning $\delta_0$ is nearly free from Zeeman shifts~\cite{Cesar96}. The $|e\rangle-|e'\rangle$ detuning $\Delta(B)\approx\Delta-\mu_{\rm B}B/\hbar$, on the other hand, is field-sensitive and has a position-dependent shift.

We consider a 2S-3P lattice composed of three pairs of standing wave Gaussian beams, each with $1/e^2$  diameter $d$, arranged symmetrically with equal intersection angles $\alpha$ to $\hat z$. The choice of relative phases between standing waves is not critical to the cooling scheme. The beams are circularly polarized to maximize the $\sigma^+$ components (relative to $\hat B$). In the Paschen-Back regime considered here, with $\mu_{\rm B}B\gg \hbar \Delta_{\rm 3P, fine}$ ($\Delta_{\rm 3P, fine}/(2\pi)= 3.25$~GHz), the $\pi$ coupling induces spin flip losses after 2S excitation with a branching ratio of $r_{\rm sf}=\frac{2}{9}\Delta_{\rm 3P, fine}^2/(\Delta +\mu_{\rm B} B/\hbar)^2$ (similar for $\sigma^-$ coupling). Even for 2S-3P light that is purely $\pi$ or $\sigma^-$ polarized, the spin-flip probability per 2S excitation is still less than $0.3\%$ in a field of 1~T.

\begin{figure}
\includegraphics [width=3.4 in,angle=0] {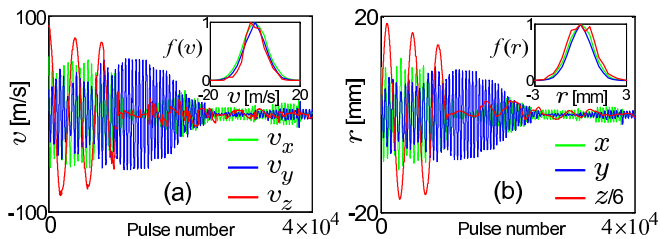}
\caption{Evolution of atomic velocity (a) and position (b) for a typical classical trajectory during the simulated cooling of magnetically trapped H. The insets give the quasi-equilibrium distributions.}\label{figSetup}
\end{figure}

Figure~\ref{figSetup} plots a typical classical trajectory of H during cooling. The simulation starts with H in $|g\rangle$ at the trap bottom ($B=0.75$~T), with initial longitudinal ($z$) and transverse ($x, y$) kinetic energy of $E_{l}=0.5$~K and $E_{t}=$0.25~K respectively. The 2-photon excitation beam overlaps with the 2S-3P beams in a 12~cm long and 1.8 cm wide cooling zone, approximately covering the trap up to the 0.2~K equipotential surface~\cite{exp:foot4}. We choose $d=3$~cm and $\alpha=0.1$ for the 2S-3P beams, with peak intensity of 0.46~kW/cm$^2$ per beam corresponding to $\Omega_{ee'}/(2\pi)=$1.3~GHz. $\Delta(0.75{\rm T})/(2\pi)=5.3$~GHz is chosen so that $\Delta(B)\gg \gamma_{e'g}$ within the cooling zone.  In a flatter octopole trap~\cite{octupoletrap} the detuning constraint is reduced, allowing for a reduced  $\Omega_{ee'}$ and less 656~nm power. The total power requirements can also be lessened with a moderate finesse optical cavity. We choose $\theta=\pi/8$, $\tau=4$~ns and $T_{\rm rep}=2$~$\mu$s~\cite{exp:foot7}. The 2-photon detuning $\delta_0=-\pi/\tau$ is chosen to improve the scattering rate for longitudinally cold atoms that are transversely hot.

The rapid cooling trajectory shown in Fig.~\ref{figSetup} is typical for atoms with $E_{l}<0.5$~K and $E_{t}<0.25$~K, which are cooled to quasi-equilibrium within $N_{\rm total}=4\times10^4$ pulses during a cooling time of only 80 ms. While some atoms with $E_{t}$ significantly larger than 0.2~K may orbit around the cooling zone and not be efficiently cooled, the final velocity distribution for most atoms is remarkably isotropic with $5$~m/s width (Fig.~\ref{figSetup}a inset), which can be further reduced by increasing $\tau$ (Fig.~\ref{FigCQ}b). The total spin-flip loss is found to be less than $0.1\%$. {{As with other hydrogen cooling proposals~\cite{Allegrini93,Zehnle01,ultrafast06}, one must consider limitations imposed by photoionization losses. We perturbatively calculate photoionization from state populations determined by internal state dynamics that ignore photoionization.}} For a reasonable two-color 2-photon excitation scheme where the stronger laser beam cannot ionize H from the 2S state in a single step~\cite{Yatsenko99}, we found ionization losses less than 25$\%$~\cite{exp:foot3}.  Even if 243~nm radiation is used for the 1S-2S excitation~\cite{Haas06}, photoionization loss would still be less than 25$\%$ for cooling, by applying $N_{\rm total}=10^9$, $\theta=2.5$~mrad pulses~\cite{exp:foot8} in $2000$~s.

We have proposed and analyzed a pulsed Sisyphus laser cooling scheme applicable to magnetically trapped H or $\overline {\rm H}$. The approach leads to rapid 3D cooling to $<$10~mK in a magnetic trap over a large volume with small spin-flip losses. Approaches to reduce photoionization losses to practical levels are proposed. Cooling efficiency may be further improved by exploring the spatial-temporal control of both the 2-photon excitation and the excited state lattice. This excited-state Sisyphus method may open new possibilities for cooling of deuterium, tritium or other species with metastable states.

\begin{acknowledgments}
We would like to thank Amy Cassidy, Gretchen Campbell, and Jonathan Wrubel for helpful discussions.
\end{acknowledgments}


\begin{thebibliography}{99}
\bibitem{exp:foot5}
Rising/falling edges are smoothed by sine functions over $20\%$ of the pulse duration.
\bibitem{exp:foot4}
To avoid excitation outside the cooling zone, one can overlap two excitation beams with distinct frequencies of $\omega_1$ and $\omega_2=\omega_{\rm 2S}-\omega_1$ respectively, where $\omega_{\rm 2S}$ is the frequency of the 1S-2S transition. Also see ref.~\cite{Yatsenko99}.
\bibitem{exp:foot7}
To allow slowest atoms confined by the $|e\rangle-|e'\rangle$ lattice to decay, the $|e\rangle \rightarrow |g\rangle$ decay rate ($\gamma_{eg}$) is increased from 0 to 10 $\mu$s$^{-1}$ for $t>1$~$\mu$s. {{This can be effectively realized with a broadband $|e\rangle-|e'\rangle$ coupling pulse.}}
\end{thebibliography}
\end{document}